\documentstyle[12pt,epsfig]{article}
\newcommand{\be}{\begin{equation}}
\newcommand{\ee}{\end{equation}}
\newcommand{\beqn}{\begin{eqnarray}}
\newcommand{\eeqn}{\end{eqnarray}}
\newcommand{\fP}{I\!\!P}
\newcommand{\spc}[1]{\mbox{\hspace{#1}}}

\newcounter{savefig}

\begin{document}
\begin{flushright} \raisebox{0.5cm}{ANL-HEP-PR-97-03} \\ 
\end{flushright}
\begin{center}
\begin{large}
Large Rapidity Gap Events in Deep Inelastic Scattering
\vspace{1cm}

M.W\"usthoff \\
\end{large}
{\it High Energy Physics Division, Argonne National Laboratory, Argonne, 
IL 60439, USA}
\\
\end{center}

\vspace{1.5cm} {\bf Abstract:} Large Rapidity Gap Events 
in Deep Inelastic Scattering are discussed in terms of 
lightcone wave functions for quarks and gluons inside the photon.
It is shown that this approach is consistent with earlier, 
conventional Feynman diagram calculations. An updated
parametrization for the cross section is given and a numerical
analysis presented. \\

\section{Introduction}
With the start of HERA a new tool has become available 
which allows a significant
progress in sorting out the longstanding puzzle about the 
nature of the Pomeron \cite{ZEUSdiff,H1diff}. Theoretically the 
Pomeron is difficult to tackle. It was introduced phenomenologically as
a simple moving pole in the complex angular momentum plane (Pomeron
trajectory) analogous to meson exchanges, however, compared to mesons there
is no clear evidence for associated bound states in the s-channel 
(candidates are glueballs). The Pomeron intercept is slightly
above one which translates into a slowly growing total and elastic 
cross section in hadron reactions \cite{DL} (soft Pomeron). Meson
trajectories, on the other hand, are below one and give subleading
contributions at very high energies. 
Within the framework of perturbative QCD the
Pomeron is associated with the resummation of leading logarithms 
in $s$ (total energy) which results in
a more complicated branch point singularity instead of a simple pole \cite{BFKL}
(hard Pomeron). The major shortcoming of this leading log($s$) approach is 
the ignorance of nonperturbative contributions and the neglect of
unitarity corrections which are relevant for 
the complete formation of the Pomeron \cite{BarWue,Bar}.
Other approaches propose a combination of soft and hard Pomeron 
(see for example \cite{Capella,ALLM}) where the Pomeron intercept is
controlled by the relevant scale of the process. 

Due to its zero color charge the Pomeron is associated with the occurrence of
rapidity gaps at very high energies. The simplest form of rapidity gap
events beside elastic scattering is single diffraction where only one of 
the incoming particles dissociates (the virtual photon in 
Deep Inelastic Scattering) whereas the other (the proton) stays intact. 
In most cases the proton escapes undetected,
but it is surrounded by a Large Rapidity Gap, a fact which is used in 
experiment to define diffraction. In the following 
we will assume the proton not to decay.
 
Deep Inelastic Scattering exhibits the nice feature of having
a small colorless particle, the virtual photon, in the intial
state. We will make use of this fact 
and shift the focus from the less well defined Pomeron to 
the virtual photon which we believe is perturbatively
calculable, i.e. its content of quarks and gluons can be determined.
Going into the target rest frame the following picture emerges:
a fast traveling photon dissociates far upstream the proton-target
into a quark-antiquark pair which evolves into a more complex partonic system
before the actual interaction takes place. 
The initial separation of the quark-antiquark pair in impact parameter 
space is of the order of $1/Q$ ($Q^2$ is the photon virtuality). The final 
partonic system, however, which takes part in the interaction
covers a much larger area roughly the size of a hadron. It appears to be
rather natural to refer to the leading order quark-antiquark pair as 
a color dipole, it is less trivial, though, for a multi particle state.
Still, it may be shown that for leading twist contributions 
the concept of a color dipole
can be extended to a more complicated final state. The argument goes as
follows: 
hard QCD-radiation generates a bunch of partons strongly ordered in
impact parameter space. The large distance is marked by 
the last quark or gluon in the chain of emissions whereas 
the remaining partons are confined in a small area (short distances).
The separated parton on the one hand and the confined system of partons
on the other hand form a new effective dipole.
Both types, the quark dipole and the gluon dipole, can be described in terms of
lightcone wave functions. The gluon dipole, although it is 
of higher order in perturbation theory, 
is of particular relevance when the invariant
mass $M$ of the diffractive final state becomes large.
The dipole picture is certainly limited in its
applicability, but it works fine when $Q^2$ is 
the leading scale in the process, i.e. $M^2/Q^2$
is not extremely large. 

For the interaction of the color dipole with the
proton we employ the two gluon model \cite{LN}. All possible 
couplings have to be added up in order to retain gauge invariance. 
The two gluon model itself may not
account for the full structure of the Pomeron, but it can 
be generalized to the exchange any number of
gluons, since all gluons hooked on to one leg of the dipole
can be merged into a single effective vertex. The strategy in this paper
is to factorize the Pomeron structure from the dipole 
according to the $k_t$-factorization theorem \cite{Cat}. 
The dipole part is calculated whereas the Pomeron part acquires 
a phenomenological parametrization. A full QCD-treatment is not feasible at the
present time. All free parameters will be determined 
from inclusive Deep Inelastic Scattering data ($F_2$) and then used
for diffraction.

In the following section we will introduce the lightcone 
wave function formalism which includes an improved expression 
for the gluon dipole.
In section 3 the model for the Pomeron is specified and a fit to $F_2$-data
which determines all parameters is performed. The Pomeron model is then combined
with the lightcone wave functions and the cross section for diffraction in Deep
Inelastic Scattering is calculated. It is shown that 
these results are consistent with
earlier approaches based on more conventional calculations using Feynman
diagrams (section 4). In section 5 the cross section is numerically
evaluated and the main results are discussed.  

\section{Lightcone Wave Functions}
One should take the terminology 'wave function' not too literally, since
the state it is meant to represent is not real but virtual. Still, 
it provides an effective and intuitive description 
of quark dipoles and gluon dipoles inside the photon in close analogy to a
quark-antiquark boundstate (quarkonium). 
An unpleasant property which 
can be traced back to the pointlike nature of the
photon is the lack of normalizability of the wave function 
unlike for real states. In order to find the 
correct normalization one has to go back
to the corresponding Feynman diagrams \cite{Mue,NZ}. 
The nice feature is that once having determined the lightcone wave
function one can easily study single gluon and multi gluon exchange,
although the wave function formalism for the gluon dipole is restricted
to color zero exchange, i.e. it works for two or multi gluon exchange.
\begin{figure}[h]
  \begin{center}
    \leavevmode
\begin{picture}(0,0)%
\epsfig{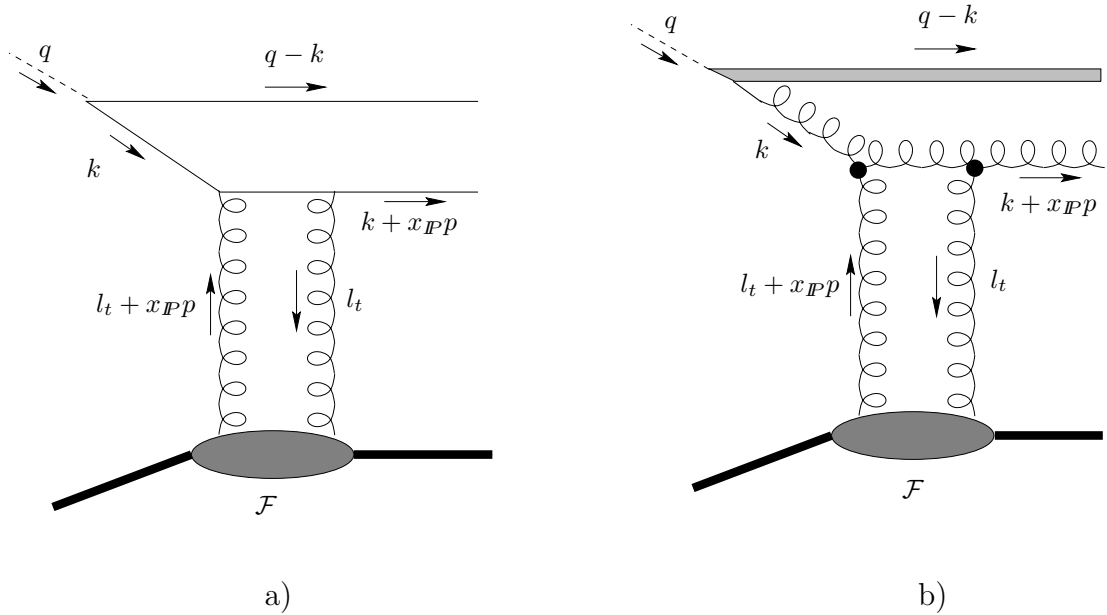}%
\end{picture}%
\setlength{\unitlength}{0.00033300in}%
\begingroup\makeatletter\ifx\SetFigFont\undefined
\def\x#1#2#3#4#5#6#7\relax{\def\x{#1#2#3#4#5#6}}%
\expandafter\x\fmtname xxxxxx\relax \def\y{splain}%
\ifx\x\y   
\gdef\SetFigFont#1#2#3{%
  \ifnum #1<17\tiny\else \ifnum #1<20\small\else
  \ifnum #1<24\normalsize\else \ifnum #1<29\large\else
  \ifnum #1<34\Large\else \ifnum #1<41\LARGE\else
     \huge\fi\fi\fi\fi\fi\fi
  \csname #3\endcsname}%
\else
\gdef\SetFigFont#1#2#3{\begingroup
  \count@#1\relax \ifnum 25<\count@\count@25\fi
  \def\x{\endgroup\@setsize\SetFigFont{#2pt}}%
  \expandafter\x
    \csname \romannumeral\the\count@ pt\expandafter\endcsname
    \csname @\romannumeral\the\count@ pt\endcsname
  \csname #3\endcsname}%
\fi
\fi\endgroup
\begin{picture}(17303,9458)(1465,-10045)
\put(15677,-887){\makebox(0,0)[lb]{\smash{\SetFigFont{10}{12.0}{rm}$q-k$}}}
\put(11718,-963){\makebox(0,0)[lb]{\smash{\SetFigFont{10}{12.0}{rm}$q$}}}
\put(5476,-9961){\makebox(0,0)[lb]{\smash{\SetFigFont{12}{14.4}{rm}a)}}}
\put(5328,-8553){\makebox(0,0)[lb]{\smash{\SetFigFont{10}{12.0}{rm}$\cal{F}$}}}
\put(13126,-2986){\makebox(0,0)[lb]{\smash{\SetFigFont{10}{12.0}{rm}$k$}}}
\put(15423,-8298){\makebox(0,0)[lb]{\smash{\SetFigFont{10}{12.0}{rm}$\cal{F}$}}}
\put(16804,-5013){\makebox(0,0)[lb]{\smash{\SetFigFont{10}{12.0}{rm}$l_t$}}}
\put(12901,-5086){\makebox(0,0)[lb]{\smash{\SetFigFont{10}{12.0}{rm}$l_t+x_{\fP}p$}}}
\put(16951,-3811){\makebox(0,0)[lb]{\smash{\SetFigFont{10}{12.0}{rm}$k+x_{\fP}p$}}}
\put(15676,-9961){\makebox(0,0)[lb]{\smash{\SetFigFont{12}{14.4}{rm}b)}}}
\put(6753,-5312){\makebox(0,0)[lb]{\smash{\SetFigFont{10}{12.0}{rm}$l_t$}}}
\put(6976,-4111){\makebox(0,0)[lb]{\smash{\SetFigFont{10}{12.0}{rm}$k+x_{\fP}p$}}}
\put(2851,-5386){\makebox(0,0)[lb]{\smash{\SetFigFont{10}{12.0}{rm}$l_t+x_{\fP}p$}}}
\put(2701,-3286){\makebox(0,0)[lb]{\smash{\SetFigFont{10}{12.0}{rm}$k$}}}
\put(5477,-1487){\makebox(0,0)[lb]{\smash{\SetFigFont{10}{12.0}{rm}$q-k$}}}
\put(1952,-1412){\makebox(0,0)[lb]{\smash{\SetFigFont{10}{12.0}{rm}$q$}}}
\end{picture}
  \end{center}
  \caption{Quark dipole (a) and gluon dipole (b)}
  \label{fig1}
\end{figure}

The virtual photon can be transverse polarized (transverse with respect to 
the lightcone vectors $q'=q+x_Bp$ and $p$, $q$ is the photon
momentum  and $p$ the momentum of the incoming proton) as well as
longitudinally polarized. The two helicity states $\gamma=\pm 1$ follow from the
projection on the transverse vectors (1,i) and (1,-i).
For the left and right handed quarks we introduce the quark helicity $h=\pm 1$. 
The quark momentum $k$ may be parametrized like $k=\alpha q'+ \beta_k p +k_t$
(Sudakov parametrization) which simultaneously fixes the decomposition of the
antiquark momentum: $q-k=(1-\alpha)q'+ (-x_B-\beta )p-k_t$. 
Using complex notation for the two dimensional vector $k_t$ we may write
the lightcone wave function for the quark-antiquark state as:
\beqn\label{lw1}
\Psi^{\gamma}_h(\alpha,k_t)
&=&\left\{ \begin{array}{ll}
\begin{displaystyle}
\frac{\sqrt{2}\;(\alpha -1)\;k_t}{|k_t|^2+\alpha (1-\alpha)Q^2}
\end{displaystyle}&\;\mbox{for $\gamma=+1$ and $h=+1$}
\\ & \\ \begin{displaystyle}
\frac{\sqrt{2}\; \alpha\;k_t}{|k_t|^2+\alpha (1-\alpha)Q^2}
\end{displaystyle}&\;\mbox{for $\gamma=+1$ and $h=-1$}
\\ & \\ \begin{displaystyle}
\frac{\sqrt{2}\; \alpha\;k_t^*}{|k_t|^2+\alpha (1-\alpha)Q^2}
\end{displaystyle}&\;\mbox{for $\gamma=-1$ and $h=+1$}
\\ & \\ \begin{displaystyle}
\frac{\sqrt{2}\;(\alpha -1)\;k_t^*}{|k_t|^2+\alpha (1-\alpha)Q^2}
\end{displaystyle}&\;\mbox{for $\gamma=-1$ and $h=-1$}
\end{array} \right. \\  \nonumber \; 
\eeqn
and
\beqn\label{lw2}
\Psi^{\gamma}_h(\alpha,k_t)&=& \spc{0.5cm} 2\;\frac{\alpha (1-\alpha) \;Q}
{|k_t|^2+\alpha (1-\alpha)Q^2} \spc{0.48cm} \mbox{for $\gamma=\ 0$ and 
$h=\pm 1$}
\eeqn
where $Q^2=-q^2$ is the virtuality of the photon. Eq.(\ref{lw2}) shows
the lightcone wave function of the longitudinally polarized photon.

Before discussing the properties of the lightcone wave function
it is necessary to have a closer look at the kinematics. We assume that
either the quark is offshell and the antiquark onshell or  vice versa
(here: $(q-k)^2=0$ and $k^2\ne 0$, fig.\ref{fig1}). In the frame that we
choose the photon moves fast and the two quarks roughly carry the
momenta $\alpha q'$ and $(1-\alpha )q'$ whereas the other components 
are small. Any subsequent high energy scattering does not change 
the $\alpha$-component. The offshell quark with the momentum $k$ 
becomes onshell after receiving a small fraction $x_{\fP}$ of momentum along 
$p$ while being scattered. The momentum of the quark changes from $k$ to
$\tilde{k}$ with $\tilde{k}^2=0$. Using the mass-shell condition one
finds:
\beqn\label{lw3}
\tilde{k}&=&\alpha q' + \frac{|k_t|^2}{\alpha W^2}p + k_t \\
q-k&=&(1-\alpha) q' +\frac{|k_t|^2}{(1-\alpha) W^2}p - k_t
\nonumber\;\;.
\eeqn
$W$ is the total hadronic mass ($W^2=2q'\cdot p$). 
The missing mass $M$ is simply given as the total energy of
the two outgoing quarks:
\be\label{lw4}
M^2\;=\;(\tilde{k}+q-k)^2=\left(q'+\frac{|k_t|^2}{\alpha (1-\alpha )
W^2}p \right)^2\;=\;\frac{|k_t|^2}{\alpha (1-\alpha )}\;\;.
\ee  

We now include the emission of a gluon in our discussion. At large
$M$ the gluon is well separated in rapidity from the $q\bar{q}$-pair 
and becomes the dominant configuration over 
the exclusive $q\bar{q}$-pair production. The latter is
suppressed by a power in $M^2$ (spin 1/2-exchange). 
Unfortunately, the three particle Fock
state is much more complicated, and a rigorous construction of the wave
function which is consistent with Feynman rules and valid for 
all kinematics has not been achieved,
yet. A simplification occurs when only the leading twist and leading
log($Q^2$) contribution is considered. In this case the distance in the
impact parameter space between the quark and the antiquark
is much smaller than the distance between the quarks and the gluon. The
$q\bar{q}$-pair on the one side and the gluon on the other side
form an effective color dipole similar to the exclusive
$q\bar{q}$-pair that we have considered before 
(fig.\ref{fig1}.b).   

For the gluon dipole we find the following wave function
(after introducing the vector notation $k_t=(k_t^1,k_t^2)$):
\be\label{lw5}
\Psi^{mn}(\alpha,k_t)\;=\;\frac{1}{\sqrt{\alpha(1-\alpha)Q^2}}\;
\frac{k_t^2\;\delta^{mn}\;-\;2\;k_t^mk_t^n}{k_t^2+\alpha(1-\alpha)Q^2}\;\;.
\ee
In view of the two-vector-particles state (an effective two gluon
state) it appears rather natural to find a tensor representation for the
wave function.
In the triple Regge limit ($TRL$) with $M^2$ much larger
than $Q^2$ the term $\alpha(1-\alpha)Q^2$ in the denominator of 
expr.(\ref{lw5}) may be neglected and only $k_t^2$ remains:
\be\label{lw5.5}
\Psi^{mn}_{TRL}(\alpha,k_t)\;=\;\frac{1}{\sqrt{\alpha(1-\alpha)Q^2}}\;
\frac{2\;k_t^mk_t^n}{k_t^2}\;\;.
\ee
The $\delta$-term which at first sight should be kept was also
removed, since it does not depend on $k_t$ and drops out 
in any application due to subtractions (see eq.(\ref{diff7}) of section
4). The simple structure of eq.(\ref{lw5.5}) 
was found earlier in refs.\cite{Mue,Rys}. Expr.(\ref{lw5}), on the
other hand, is valid for all masses and provides the natural 
extension of the Triple Regge result. 
 
We have to point out that the wave function introduced in
eq.(\ref{lw5}) does not reproduce the amplitude for the single gluon
t-channel exchange. Especially those contributions which are singular 
in the limit
$M\rightarrow 0$ are absent. The reason for that lies in the fact that 
in the color singlet configuration all singular, soft terms cancel out. 
This cancellation can be illustrated by considering
final state radiation off the $q\bar{q}$-pair. The soft terms add up
when the $q\bar{q}$-pair is colored, they cancel each other, however, when
the state is colorless.

We have written eq.(\ref{lw5}) in a symmetric way with respect to
$\alpha$ and $1-\alpha$ in order to stress the similarity with the
$q\bar{q}$-dipole. The gluon dipole considered here is actually very
asymmetric which is a consequence of the leading twist
approximation where the internal virtualities are much smaller than
$Q^2$. From eq.(\ref{lw3}) we conclude that $\alpha$ has to be much
smaller than 1 in order to fulfill the condition $k^2\ll Q^2$. So, one
could have set $1-\alpha$ in eq.(\ref{lw5}) equal to 1 without
reducing the accuracy of the formula.

At very large masses $M$ or small $\beta$ the dipole picture becomes
insufficient. Instead of logarithms in $Q^2$ we have to sum
up logs in $1/\beta$ or $M^2/Q^2$. This leads to a new
four-gluon t-channel state \cite{BarWue,BLW} and a new 
singularity $1+\omega_4$ in the complex angular momentum plane 
($(1/\beta)^{\omega_4}$). But in contrast to conventional 
QCD-scaling violation the strong rise
with $1/\beta$ does not imply a rise with $Q^2$.

\section{Modeling the Structure Function $F_2$}

We compute the structure function $F_2$ at very low $x_B$ when single 
gluon exchange gives the leading contribution. In contrast to the
usual approach (gluon-boson fusion) the gluon is not onshell and
needs to be described by a distribution over both, longitudinal and 
transverse, phase space components ($k_t$-factorization). 
Such a factorized distribution ($\cal{F}$) contains perturbative as well
as nonperturbative contributions,
and the aim is to find a suitable parametrization which gives a
reasonable description of all low $x_B$ and low $Q^2$ data.
\begin{figure}[h]
  \begin{center}
    \leavevmode
\begin{picture}(0,0)%
\epsfig{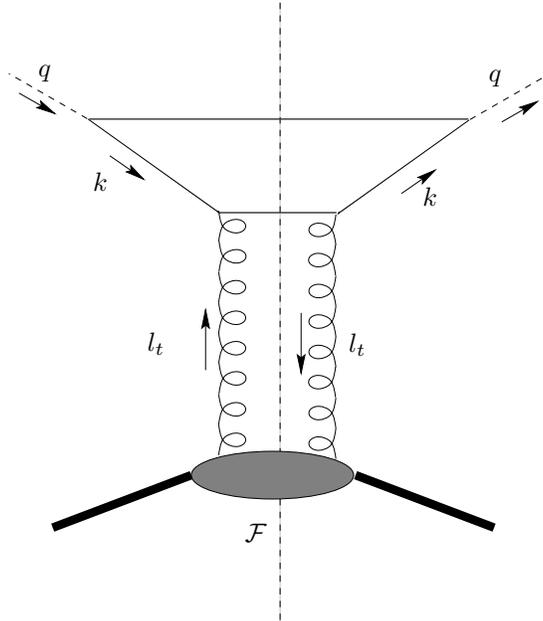}%
\end{picture}%
\setlength{\unitlength}{0.00033300in}%
\begingroup\makeatletter\ifx\SetFigFont\undefined
\def\x#1#2#3#4#5#6#7\relax{\def\x{#1#2#3#4#5#6}}%
\expandafter\x\fmtname xxxxxx\relax \def\y{splain}%
\ifx\x\y   
\gdef\SetFigFont#1#2#3{%
  \ifnum #1<17\tiny\else \ifnum #1<20\small\else
  \ifnum #1<24\normalsize\else \ifnum #1<29\large\else
  \ifnum #1<34\Large\else \ifnum #1<41\LARGE\else
     \huge\fi\fi\fi\fi\fi\fi
  \csname #3\endcsname}%
\else
\gdef\SetFigFont#1#2#3{\begingroup
  \count@#1\relax \ifnum 25<\count@\count@25\fi
  \def\x{\endgroup\@setsize\SetFigFont{#2pt}}%
  \expandafter\x
    \csname \romannumeral\the\count@ pt\expandafter\endcsname
    \csname @\romannumeral\the\count@ pt\endcsname
  \csname #3\endcsname}%
\fi
\fi\endgroup
\begin{picture}(8516,9794)(1465,-9983)
\put(6798,-5717){\makebox(0,0)[lb]{\smash{\SetFigFont{10}{12.0}{rm}$l_t$}}}
\put(7951,-3436){\makebox(0,0)[lb]{\smash{\SetFigFont{10}{12.0}{rm}$k$}}}
\put(8984,-1501){\makebox(0,0)[lb]{\smash{\SetFigFont{10}{12.0}{rm}$q$}}}
\put(5176,-8686){\makebox(0,0)[lb]{\smash{\SetFigFont{10}{12.0}{rm}$\cal{F}$}}}
\put(3655,-5701){\makebox(0,0)[lb]{\smash{\SetFigFont{10}{12.0}{rm}$l_t$}}}
\put(1952,-1412){\makebox(0,0)[lb]{\smash{\SetFigFont{10}{12.0}{rm}$q$}}}
\put(2806,-3196){\makebox(0,0)[lb]{\smash{\SetFigFont{10}{12.0}{rm}$k$}}}
\end{picture}
  \end{center}
  \caption{Inclusive Deep Inelastic Scattering ($F_2$).}
  \label{fig2}
\end{figure}

The $k_t$-factorization theorem is the high energy or small $x_B$
counterpart of the conventional (collinear) factorization theorem.
The latter, when calculating the structure function $F_2$,
requires a convolution of the gluon structure function and the quark box
with respect to the longitudinal momentum fraction whereas
in the small $x_B$ regime the convolution with respect to the
transverse momentum ($k_t$) is more appropriate. At zero momentum transfer 
the Pomeron (in Deep Inelastic Scattering) is essentially the same as the 
unintegrated gluon structure function, and by fitting the $F_2$-data 
one determines the Pomeron intercept $\alpha_{\fP}$. In Deep Inelastic
Scattering, however, the Pomeron intercept varies (depending on $Q^2$) 
rather than being a fixed number as in soft processes
($\alpha_{\fP}=1.085$).

As we already mentioned one of the virtues of the wave function
formalism is that we can use it for single gluon exchange 
(transverse momentum $l_t$) as well (see fig.\ref{fig2}). We find for
$F_2$ the following expression:
\beqn\label{f2_1}
F_2(x_B,Q^2)&=&\sum_f Q_f^2\;\frac{Q^2}{4\pi}\;\int \frac{d^2l_t}{\pi\;l_t^2}
\;{\cal F}(x_B,l_t^2,Q_0^2)\;\;\cdot\\ &&\cdot\;\;\nonumber
\int_0^1 d\alpha\;\int \frac{d^2k_t}{4 \pi}\;\sum_{\gamma=0,+,-}\;\sum_{h=+,-}\;
\left|\Psi^{\gamma}_h(\alpha,k_t)\;-\;\Psi^{\gamma}_h(\alpha,k_t+l_t)
\right|^2\;\;.
\eeqn
In the limit of large $Q^2$ (at fixed but small $x_B$) and taking 
only the leading log($Q^2$) contribution the gluon distribution factorizes
and the $l_t$-integral can be taken: 
\be\label{f2_2}
x_B g(x_B,Q^2)\;=\;\int_0^{Q^2}dl_t^2\;\frac{1}{\alpha_s}
\;{\cal F}(x_B,l_t^2,Q_0^2)\;\;.
\ee
$g(x_B,Q^2)$ represents the conventional gluon distribution and
${\cal F}/\alpha_s$ is usually referred to as unintegrated gluon structure
function. Introducing the Feynman parameter $x$ eq.(\ref{f2_1}) reduces to
\beqn\label{f2_3}
F_2(x_B,Q^2)&=&\sum_f Q_f^2\;\frac{Q^2}{4\pi}\;\int dl_t^2
\;{\cal F}(x_B,l_t^2,Q_0^2)\;\;\cdot\\ &&\cdot\;\;\nonumber
\int_0^1 dx\int_0^1 d\alpha \;\frac{[1-2x(1-x)][1-2\alpha(1-\alpha)]
+8x(1-x)\alpha(1-\alpha)}{x(1-x)l_t^2+\alpha(1-\alpha)Q^2}\;\;.
\eeqn
This representation (see also \cite{LevRys,KMRS}) serves as starting
point for further numerical evaluation.

As ansatz for ${\cal F}$ we choose:
\beqn\label{f2_4}
{\cal F}(x_B,l_t^2,Q_0^2)&=&\frac{G(x_B,Q^2/Q_0^2)}{l_t^2+Q_0^2}\\
G(x_B,Q^2/Q_0^2)&=&A\;\;\left(\frac{x_0}{x_B}\right)
^{1-\alpha_{\fP}(Q^2)}\;
\left[\ln\left(\frac{Q^2}{Q_0^2}\right)+1\right]^{-C}\nonumber
\eeqn
$Q_0$ is set to $1 GeV$ (proton mass), and since only small $x_B$ 
are considered we introduced $x_0=0.05$ as normalization point. The 
Pomeron intercept has the following parametrization:
\be\label{f2_5}
\alpha_{\fP}(Q^2)\;=\;0.085\;+\;\left\{
\begin{array}{ll} B\;\ln[\ln(Q^2/Q_0^2)+1] &\mbox{if 
\hspace{0.5cm} $Q^2 > Q_0^2$}
\\ 0 &\mbox{if \hspace{0.5cm} $Q^2 \leq Q_0^2$}  \end{array} \right.
\ee
A soft Pomeron ($\alpha_{\fP}=1.085$) intercept is assumed 
when the scale $Q^2$ falls below $Q_0^2$. This behavior has experimental
support form the BPC-data \cite{ZEUSBPC}

The ansatz in eq.(\ref{f2_4}) has the following two basic ingredients: 
first, ${\cal F}$ scales like $1/Q_0^2$ for $l_t=0$, i.e. an effective cutoff at
the scale of $Q_0^2$ is introduced which eliminates the singularity related
to the gluon propagator $1/l_t^2$. The scale $Q_0$ roughly represents the 
inverse size of the hadron and because hadrons are colorless all gluons
with a wave length larger than the size of the hadron decouple. One
important consequence is the vanishing of the structure function when $Q^2$
approaches zero. The second ingredient is a scale ($Q^2$-) dependent
Pomeron intercept. It takes care of the experimental fact that the
small-$x_B$ rise becomes weaker when the scale decreases and  
and finally turns into the soft behavior below $Q_0$. The parameter
$B$ in (\ref{f2_5}) has to be determined from data. Two more parameters
come along with $B$ (see eq.(\ref{f2_4})), 
$A$ fixes the absolute normalization and 
$C$ corrects the strong scaling violation which results from the scale
dependent parametrization of the Pomeron intercept.

The data for the fit are taken from HERA  \cite{ZEUSBPC,H1f2}
below $Q^2=50GeV^2$ and from E665 \cite{E665} including
the smallest $Q^2$-values. The fit gives the following
values for the three parameters $A, B$ and $C$:
\beqn\label{f2_6}
A &=& 0.877 \nonumber \\
B &=& 0.133 \\
C &=& 0.596 \nonumber
\eeqn

These parameters will be used in the following to predict the
diffractive structure function $F_2^D$.
\section{Diffraction} 
For the diffractive cross section we use the same conventions as 
\begin{figure}[h]
  \begin{center}
    \leavevmode
\begin{picture}(0,0)%
\epsfig{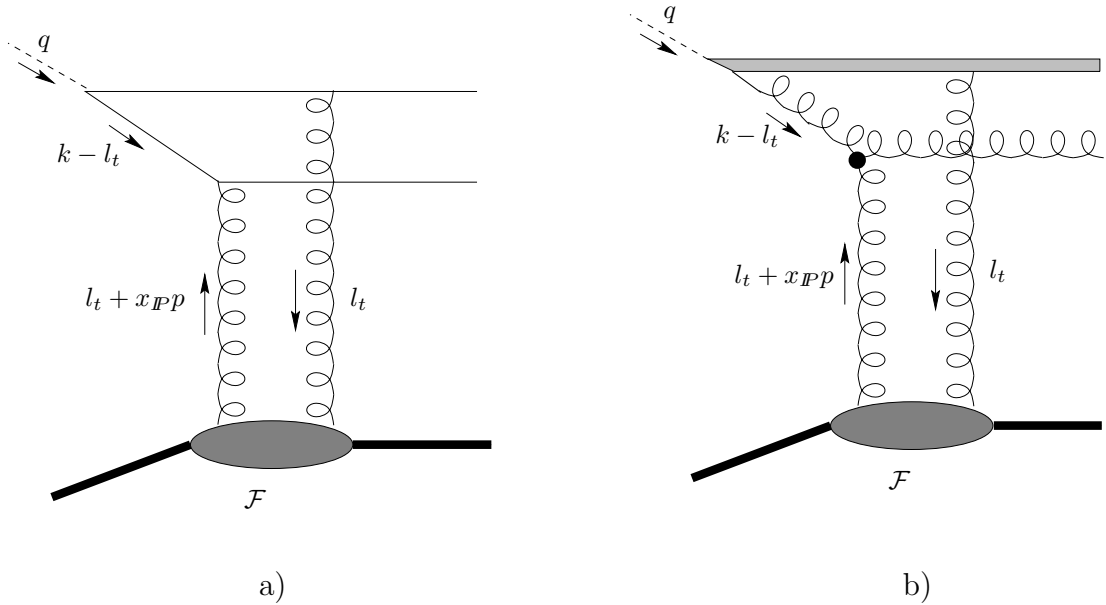}%
\end{picture}%
\setlength{\unitlength}{0.00033300in}%
\begingroup\makeatletter\ifx\SetFigFont\undefined
\def\x#1#2#3#4#5#6#7\relax{\def\x{#1#2#3#4#5#6}}%
\expandafter\x\fmtname xxxxxx\relax \def\y{splain}%
\ifx\x\y   
\gdef\SetFigFont#1#2#3{%
  \ifnum #1<17\tiny\else \ifnum #1<20\small\else
  \ifnum #1<24\normalsize\else \ifnum #1<29\large\else
  \ifnum #1<34\Large\else \ifnum #1<41\LARGE\else
     \huge\fi\fi\fi\fi\fi\fi
  \csname #3\endcsname}%
\else
\gdef\SetFigFont#1#2#3{\begingroup
  \count@#1\relax \ifnum 25<\count@\count@25\fi
  \def\x{\endgroup\@setsize\SetFigFont{#2pt}}%
  \expandafter\x
    \csname \romannumeral\the\count@ pt\expandafter\endcsname
    \csname @\romannumeral\the\count@ pt\endcsname
  \csname #3\endcsname}%
\fi
\fi\endgroup
\begin{picture}(17303,9382)(1465,-10045)
\put(11718,-963){\makebox(0,0)[lb]{\smash{\SetFigFont{10}{12.0}{rm}$q$}}}
\put(16804,-5013){\makebox(0,0)[lb]{\smash{\SetFigFont{10}{12.0}{rm}$l_t$}}}
\put(15226,-8311){\makebox(0,0)[lb]{\smash{\SetFigFont{10}{12.0}{rm}$\cal{F}$}}}
\put(15451,-9961){\makebox(0,0)[lb]{\smash{\SetFigFont{12}{14.4}{rm}b)}}}
\put(12826,-5086){\makebox(0,0)[lb]{\smash{\SetFigFont{10}{12.0}{rm}$l_t+x_{\fP}p$}}}
\put(5176,-8611){\makebox(0,0)[lb]{\smash{\SetFigFont{10}{12.0}{rm}$\cal{F}$}}}
\put(12526,-2911){\makebox(0,0)[lb]{\smash{\SetFigFont{10}{12.0}{rm}$k-l_t$}}}
\put(2251,-3211){\makebox(0,0)[lb]{\smash{\SetFigFont{10}{12.0}{rm}$k-l_t$}}}
\put(2701,-5461){\makebox(0,0)[lb]{\smash{\SetFigFont{10}{12.0}{rm}$l_t+x_{\fP}p$}}}
\put(1952,-1412){\makebox(0,0)[lb]{\smash{\SetFigFont{10}{12.0}{rm}$q$}}}
\put(6826,-5461){\makebox(0,0)[lb]{\smash{\SetFigFont{10}{12.0}{rm}$l_t$}}}
\put(5401,-9961){\makebox(0,0)[lb]{\smash{\SetFigFont{12}{14.4}{rm}a)}}}
\end{picture}
  \end{center}
  \caption{Wave functions with shifted argument.}
  \label{fig3}
\end{figure}
for the inclusive cross section in Deep Inelastic Scattering, i.e.
we decompose it into a transverse and a longitudinal part
according to the different polarizations of the virtual photon:
\be\label{diff1}
\left.\frac{d\sigma}{d\beta dQ^2 dx_{\fP} dt}\right|_{t=0}\;=\;
\frac{\alpha_{em}}{2 x_{\fP}
Q^4}\left\{-\;[1+(1-y)^2]x_BW_t\;+\;4(1-y)x_BW_l
\right\}
\ee
where $W_t$ and $W_l$ are the transverse and longitudinal projection
of the hadronic tensor $W^{\mu \nu}$:
\beqn\label{diff2}
W_t&=&g_t^{\mu \nu}W_{\mu \nu} \nonumber\\
W_l&=&\frac{4Q^2}{s^2}p^{\mu} p^{\nu}W_{\mu \nu}
\eeqn
with $s=2q'\cdot p$ and $q'=q+x_Bp$. The transverse tensor $g_t^{\mu
\nu}$ is defined as 
\be\label{diff3}
g_t^{\mu \nu}\;=\;g^{\mu \nu}\;-\;\frac{q'^{\mu}p^{\nu}\;+\;p^{\mu}
q'^{\nu}}{q'\cdot p}\;\;.
\ee 
The momentum transfer is set to zero, a restriction,
which is justified by the fact that the cross section peaks at $t=0$.

At leading order we have to consider the coupling of two t-channel
gluons to a quark-antiquark pair. The four possible couplings are
represented by four terms involving the wave function with shifted 
transverse momenta (see fig.\ref{fig3}.a). These shifts correspond to
the transverse momenta carried by
the gluons ($l_t$ and $-l_t$) when the total momentum transfer is zero:
\beqn\label{diff4}
&&\int \frac{d^2l_t}{\pi l_t^2}\; {\cal F}(x_{\fP},l_t^2,Q_0^2)\;\left[
2\;\Psi_h^\gamma (\alpha,k_t)\;-\;\Psi_h^\gamma (\alpha ,k_t+l_t)\;
-\;\Psi_h^\gamma (\alpha,k_t-l_t)\right] \nonumber\\
&=&\int \frac{dl_t^2}{l_t^2}\; {\cal F}(x_{\fP},l_t^2,Q_0^2)\;
\;\cdot \\
&\cdot&\left\{ \begin{array}{ll} \begin{array}{ll}
\begin{displaystyle}
\left[ \frac{k_t^2-\alpha(1-\alpha)Q^2}{k_t^2+\alpha(1-\alpha)Q^2}
\right. \end{displaystyle}\\ \begin{displaystyle}\left.
-\frac{k_t^2-l_t^2-\alpha(1-\alpha)Q^2}
{\sqrt{[k_t^2+l_t^2+\alpha(1-\alpha)Q^2]^2-4 k_t^2 l_t^2}}\right]
\end{displaystyle}\end{array}
\left\{ \begin{array}{ll}\begin{displaystyle}
\frac{\sqrt{2}\;(\alpha -1)\;k_t}{|k_t|^2}
\end{displaystyle}&\;\mbox{for $\gamma=+1$ and $h=+1$}
\\ & \\ \begin{displaystyle}
\frac{\sqrt{2}\; \alpha\;k_t}{|k_t|^2}
\end{displaystyle}&\;\mbox{for $\gamma=+1$ and $h=-1$}
\\ & \\ \begin{displaystyle}
\frac{\sqrt{2}\; \alpha\;k_t^*}{|k_t|^2}
\end{displaystyle}&\;\mbox{for $\gamma=-1$ and $h=+1$}
\\ & \\ \begin{displaystyle}
\frac{\sqrt{2}\;(\alpha -1)\;k_t^*}{|k_t|^2}
\end{displaystyle}&\;\mbox{for $\gamma=-1$ and $h=-1$}
\end{array}\right.
\\ & \\ \begin{displaystyle}
\left[ \frac{2\; \alpha (1-\alpha) \;Q}{|k_t|^2+\alpha (1-\alpha)Q^2}
-\frac{2\;\alpha(1-\alpha)\;Q}
{\sqrt{[k_t^2+l_t^2+\alpha(1-\alpha)Q^2]^2-4 k_t^2 l_t^2}}\right]
\end{displaystyle}
\;\mbox{for $\gamma=\;\; 0$ and $h=\pm 1$}
\end{array} \right. \nonumber
\eeqn
$ $\\
The last expression results after integration over the azimuthal angle.

For the total contribution to $W_t$ and $W_l$ one has to take the square of the
amplitude, i.e. basically the square of the previous expression with
summation over the corresponding helicities:
\beqn\label{diff5}
x_BW_t^{q\bar{q}}&=&-\; \sum_f Q^2_f \;\frac{\pi}{24}
\;\frac{Q^2}{\beta(1-\beta)}\;\int_0^1d\alpha\;
[\alpha^2+(1-\alpha)^2] \;
\left\{\int\frac{dl_t^2}{l_t^2}\; {\cal F}(x_{\fP},l_t^2,Q_0^2)\;\;\cdot
\right. \\
&&\left. \cdot\;\;\left[1-2\beta\;+\;
\frac{\beta \;l_t^2/Q^2\;-\;(1-2\beta)\alpha(1-\alpha)}
{\sqrt{[\beta \;l_t^2/Q^2+\alpha(1-\alpha)]^2
-4\alpha(1-\alpha)\beta(1-\beta)\;l_t^2/Q^2}} \right]\right\}^2 \nonumber
\eeqn
and
\beqn\label{diff6}
x_BW_l^{q\bar{q}}&=& \sum_f Q^2_f \;\frac{\pi}{3}
\;Q^2\;\int_0^1d\alpha\;
\alpha(1-\alpha) \;
\left\{\int\frac{dl_t^2}{l_t^2}\; {\cal F}(x_{\fP},l_t^2,Q_0^2)\;\;\cdot 
\right.\\
&&\left.
\cdot\;\;\left[1\;-\;\frac{\alpha(1-\alpha)}
{\sqrt{[\beta \;l_t^2/Q^2+\alpha(1-\alpha)]^2
-4\alpha(1-\alpha)\beta(1-\beta)\;l_t^2/Q^2}} \right]\right\}^2
\nonumber\;\;.
\eeqn
We have here substituted $k_t^2$ using eq.(\ref{lw4}) in combination
with $\beta=Q^2/(M^2+Q^2)$.
Similar expressions can be found in \cite{BarWue,Mue,NZ}. As in the
previous section we have absorbed the strong coupling constant into
$\cal{F}$. This way the remaining expressions become free of parameters. 

By taking the limits $\beta \rightarrow 1$ and $\beta \rightarrow 0$
one can study the main properties of eqs.(\ref{diff5}) and (\ref{diff6}).
For $\beta \rightarrow 1$ eq.(\ref{diff5}) vanishes proportional to
$(1-\beta)$ whereas eq.(\ref{diff6}) gives a finite contribution.
Hence, at small masses the longitudinal contribution is larger
than the transverse contribution. A similar observation is made for vector
meson production in Deep Inelastic Scattering. 
One, however, has to be aware that the longitudinal
part is of higher twist, so that the ratio of the transverse part 
to longitudinal does not vanish like $M^2/Q^2$ at large $Q^2$,
but is roughly of the order of $M^2/Q_0^2$ with a logarithmic
enhancement (see also ref.\cite{MRT}).
Taking the second limit, $\beta \rightarrow 0$, one observes that the
transverse contribution is finite, i.e. in terms of the mass $M$:
the cross section vanishes like $1/M^4$ as expected for a spin 1/2-exchange.
The longitudinal part (\ref{diff6}) has an asymptotic behaviour
proportional to $\beta^2$, i.e. is negligible at large masses.

For the configuration with a gluon in the final state we go back to 
eq.(\ref{lw5}). As in eq.(\ref{diff4}) we encounter 
four terms (see fig.\ref{fig3}.b):
\beqn\label{diff7}
&&\int \frac{d^2l_t}{\pi l_t^2}\; {\cal F}(x_{\fP},l_t^2,Q_0^2)\;\left[
2\;\Psi^{mn}(\alpha,k_t)\;-\;\Psi^{mn}(\alpha,k_t+l_t)\;-\;
\Psi^{mn}(\alpha,k_t-l_t)\right]\nonumber\\
&=&\int \frac{dl_t^2}{l_t^2}\; {\cal F}(x_{\fP},l_t^2,Q_0^2)\;
\frac{1}{\sqrt{\alpha(1-\alpha)Q^2}}
\left[1 - \frac{2k_t^2}{k_t^2+\alpha(1-\alpha)Q^2} -
\frac{l_t^2}{k_t^2} - \frac{\alpha(1-\alpha)Q^2}{k_t^2}\right.\\
&& \left.\spc{3cm}+\; 
\frac{[l_t^2-k_t^2+\alpha(1-\alpha)Q^2]^2+2k_t^2\alpha(1-\alpha)Q^2}{k_t^2
\sqrt{[l_t^2+k_t^2+\alpha(1-\alpha)Q^2]^2 -4l_t^2k_t^2}}\right]
\left\{\frac{2 k_t^m k_t^n}{k_t^2} - \delta^{mn}\right\}\nonumber\;\;.
\eeqn
To get the complete contribution for $W_t$ 
one has to, first, take the square of the previous
expression and, second, add the contribution
for the perturbative splitting of a gluon into two quarks:
\beqn\label{diff8}
&&x_BW_t^g\;=\;-\sum_f Q_f^2 \;\frac{\pi}{2}\;\int_0^{Q^2} dk^2\;
\frac{\alpha_s}{8\pi}\;\ln\left(\frac{Q^2}{k^2}\right)
\;\int_\beta^1 \frac{dz}{z^2}\;\left[\left(1-\frac{\beta}{z}\right)^2
\;+\;\left(\frac{\beta}{z}\right)^2\right]
\;\frac{9}{4}\frac{1}{(1-z)^2}\;\cdot \nonumber\\
&&\cdot\;\;\left\{\int\frac{dl_t^2}{l_t^2}
\;{\cal F}(x_P,l_t^2,Q_0^2)\;\left[z^2+(1-z)^2+\frac{l_t^2}{k^2}-
\frac{[(1-2z)k^2-l_t^2]^2+2z(1-z)k^4}{k^2\sqrt{(k^2+l_t^2)^2-4(1-z)l_t^2k^2}}
\right]\right\}^2 .
\eeqn
Two new variables were introduced in this equation, the longitudinal momentum
fraction $z$ (relative to the Pomeron momentum)
and the virtuality $k^2$ of the t-channel gluon which is
connected to the quark-box. $k^2$ is
related to the transverse momentum by $k^2=k_t^2/(1-z)$. The previously
used variable $\alpha$ was substituted by means of the equation
$\alpha(1-\alpha)Q^2=zk^2$ where we assume that $\alpha \ll 1$ according to the
leading log($Q^2$) approach adopted here. A factor of two arises in the
integration from the opposite and symmetric limit $(1-\alpha)\ll 1$.
A contribution to $W_l$ is negligible in this case.
We, again, take the two limits
$\beta \rightarrow 0$ and $\beta \rightarrow 1$ in order to understand
the basic behavior. For small $\beta$ the region of small $z$ dominates,
so that the second line in eq.(\ref{diff8})
can be approximated by setting $z$ to zero.
Integrating over $z$ then results in $1/\beta$, i.e.
the cross section is divergent when $\beta$ approaches zero. The small
$\beta$ or triple Regge limit has already been consider before in
refs.\cite{Bar,Mue,NZ,Rys,GLR}, and the results are found to be consistent
with our calculations. One can also start with
$\Psi^{mn}_{TRL}$ (eq.(\ref{lw5.5})) which directly yields 
the triple Regge limit of eq.(\ref{diff8}).
Taking the opposite limit $\beta \rightarrow 1$ which was not covered by
previous calculations one finds that eq.(\ref{diff8}) vanishes like
$(1-\beta)^3$. This result emerges by taking the limit $z \rightarrow 1$
resulting in $(1-z)^2$ and integrating over $z$. A derivation of
eq.(\ref{diff8}) based on Feynman diagrams is given in ref.\cite{Wu}.

It is important to note that the function ${\cal F}$ is assumed to be universal
and should be the same for all three eqs.(\ref{diff5}), (\ref{diff6})
and (\ref{diff8}). The only
parameter that enters in eq.(\ref{diff8}) is the strong coupling constant
$\alpha_s$. It depends on the scale somewhere between $Q_0$ and $Q$.

One may introduce a simplification by taking the limit $l_t \rightarrow
0$ in the wave functions of eqs.(\ref{diff4}) and (\ref{diff7}).
This limit gives the leading contribution provided that $k_t^2/(1-\beta)$ is
much larger than $Q_0^2$:
\beqn\label{diff9}
&&\int \frac{d^2l_t}{\pi l_t^2}\; {\cal F}(x_{\fP},l_t^2,Q_0^2)\;\left[
2\;\Psi_h^\gamma (\alpha,k_t)\;-\;\Psi_h^\gamma (\alpha ,k_t+l_t)\;
-\;\Psi_h^\gamma (\alpha,k_t-l_t)\right]\\
&\simeq& 
\left[-2 \;\frac{\partial^2 \Psi_h^\gamma (\alpha,k_t)}
{\partial k_t \partial k_t^*} \right]\;\;
\int^{|k_t|^2+\alpha(1-\alpha)Q^2}
 dl_t^2\; {\cal F}(x_{\fP},l_t^2,Q_0^2)\nonumber\\
&=& \alpha_s \;x_{\fP}
g(x_{\fP},|k_t|^2+\alpha(1-\alpha)Q^2,Q_0^2)\;\;\cdot \nonumber\\
&&\spc{1cm}\cdot\;\left\{ \begin{array}{ll}
\begin{displaystyle}
\frac{4 \alpha(1-\alpha)Q^2}{[|k_t|^2+\alpha(1-\alpha)Q^2]^3}
\end{displaystyle}
\left\{ \begin{array}{ll}\begin{displaystyle}
\sqrt{2}\;(\alpha -1)\;k_t
\end{displaystyle}&\;\mbox{for $\gamma=+1$ and $h=+1$}
\\ & \\ \begin{displaystyle}
\sqrt{2}\; \alpha\;k_t
\end{displaystyle}&\;\mbox{for $\gamma=+1$ and $h=-1$}
\\ & \\ \begin{displaystyle}
\sqrt{2}\; \alpha\;k_t^*
\end{displaystyle}&\;\mbox{for $\gamma=-1$ and $h=+1$}
\\ & \\ \begin{displaystyle}
\sqrt{2}\;(\alpha -1)\;k_t^*
\end{displaystyle}&\;\mbox{for $\gamma=-1$ and $h=-1$}
\end{array}\right.
\\ & \\ \begin{displaystyle}
\frac{\alpha(1-\alpha)Q^2-|k_t|^2}{[|k_t|^2+\alpha (1-\alpha)Q^2]^3}\;
\spc{.5cm}2\; \alpha (1-\alpha)\;Q
\end{displaystyle}
\spc{.5cm}\;\mbox{for $\gamma=\;\; 0$ and $h=\pm 1$}
\end{array} \right. \nonumber
\eeqn
One may call this result leading log($k^2/Q_0^2$) approach for
the remaining integration over $l_t^2$ (eq.(\ref{f2_2})) is logarithmic 
(one propagator $1/l_t^2$ is hidden in ${\cal F}$). It also indicates
the limitation to large $k^2$.

We again convert the previous results into contributions
to $W_t$ and $W_l$. Substituting $\alpha$ by means of
eq.(\ref{lw4}) and introducing the virtuality
$k^2=k_t^2/(1-\beta)$ we find:
\be\label{diff10}
x_BW_t^{q\bar{q}}\;=\;-\; \sum_f Q^2_f \;\frac{\pi}{3}\;
4\;\beta^2\;(1-\beta)\;\int_{k_0^2}^{Q^2} \frac{dk^2}{k^4} \;
\left\{\alpha_s \;x_{\fP}g(x_{\fP},k^2,Q_0^2)\right\}^2
\ee
and
\be\label{diff11}
x_BW_l^{q\bar{q}}\;=\; \sum_f Q^2_f \;\frac{\pi}{3 Q^2}\;
\beta^2\;(1-2\beta)^2\;\int_{k_0^2}^{Q^2} \frac{dk^2}{k^2} \;
\left\{\alpha_s \;x_{\fP}g(x_{\fP},k^2,Q_0^2)\right\}^2\;\;.
\ee

Eq.(\ref{diff10}) was derived earlier in refs.\cite{RysHERA,LeWu,dipl,NZjets} 
and eq.(\ref{diff11}) in ref.\cite{NZjets}. The virtue of this approach is
the fact that we do not need the unintegrated structure
function, instead we can use the conventional structure function.
The shortcomings, however, 
are first of all the need of a cutoff $k_0^2$ which
can only be realized by requiring jets in the final state. Second,
one recognizes that in (\ref{diff10}) the small $\beta$ region is
strongly suppressed and not constant as anticipated earlier. In this
regime next-to-leading log($k^2/Q_0^2$) corrections become important.
Corrections of this type have been explicitly calculated in \cite{BLW2}.
We face a similar situation for the longitudinal contribution
eq.(\ref{diff11}) which is zero at $\beta=1/2$. Again next-to-leading
log contributions become relevant here.

To complete the discussion on the leading log($k^2/Q_0^2$) approach
we give the corresponding formula for gluon production:
\beqn\label{diff12}
&&\int \frac{d^2l_t}{\pi l_t^2}\; {\cal F}(x_{\fP},l_t^2,Q_0^2)\;\left[
2\;\Psi^{mn} (\alpha,k_t)\;-\;\Psi^{mn} (\alpha ,k_t+l_t)\;
-\;\Psi^{mn} (\alpha,k_t-l_t)\right] \nonumber\\
&\simeq&
\left[- \;\delta^{ij}\;\; \frac{\partial^2 \Psi^{mn} (\alpha,k_t)}
{\partial k_t^i \partial k_t^j} \right]\;\; 
 \int^{|k_t|^2+\alpha(1-\alpha)Q^2}
 dl_t^2\; {\cal F}(x_{\fP},l_t^2,Q_0^2)\\
&=&
\frac{2\;k_t^2}{\sqrt{\alpha(1-\alpha) Q^2}}
\;\frac{3 \alpha(1-\alpha)Q^2 +k_t^2}
{[k_t^2+\alpha(1-\alpha)Q^2]^3}
\left\{\delta^{mn}-\frac{2 k_t^m k_t^n}{k_t^2} \right\}\;\;
\alpha_s \;x_{\fP}
g(x_{\fP},|k_t|^2+\alpha(1-\alpha)Q^2,Q_0^2) \nonumber
\eeqn
As before we take the square of the previous expression and rewrite the
result in terms of $W_t^g$. The procedure is similar to the derivation
of eq.(\ref{diff8}) where $\alpha(1-\alpha)Q^2$ is substituted by $z
k^2$ and the splitting function for gluons into quarks is added:
\beqn\label{diff13}
x_BW_t^g&=&-\sum_f Q_f^2 \;\frac{\pi}{2}\;\int_{k^2_0}^{Q^2}
\frac{dk^2}{k^4}\;
\frac{\alpha_s}{8\pi}\;\ln\left(\frac{Q^2}{k^2}\right)
\;\int_\beta^1 \frac{dz}{z^2}\;\left[\left(1-\frac{\beta}{z}\right)^2
\;+\;\left(\frac{\beta}{z}\right)^2\right]
\;\;\cdot \nonumber\\
&&\hspace{2cm}\cdot\;\;9\;(1+2z)^2\;(1-z)^2\;
\left\{\alpha_s \;x_{\fP}g(x_{\fP},k^2,Q_0^2)\right\}^2\;\;.
\eeqn
This result has been derived earlier in refs.\cite{Wu,LeWu,dipl} by 
direct calculation of Feynman diagrams. 

\section{Numerical Results}
With the formulae for Diffraction at hand and $\cal{F}$ determined by
the inclusive $F_2$ (see section 2) we can numerically evaluate
$F_2^D$. To this end we note that $F_2^D$ and $F_l^D$ are 
related to $W_t$ and $W_l$ by
\beqn\label{numeric1}
F_2^D&=&\frac{\beta}{x_{\fP}}\;
\left(-\,\frac{x_bW_t}{4\pi}\;+\;\frac{x_bW_l}{2\pi}\right) \\
F_l^D&=&\frac{\beta}{x_{\fP}}\;\frac{W_l}{4\pi} \nonumber
\eeqn
which follows from 
\be\label{numeric}
\frac{d \sigma^D}{d\beta dQ^2 dx_p}\;=\;\frac{2\pi\alpha_{em}^2}{\beta
Q^4}\; \left\{\left[1+(1-y)^2\right]\,F_2^D(\beta,Q^2,x_{\fP})
-2 x_b y^2\,F_l^D(\beta,Q^2,x_{\fP})\right\}\;\;.
\ee
The longitudinal structure function $F_l$ is usually negligible due to 
the accompanying factor $y^2$ which is experimentally small in most cases
(y is the energy loss of the electron). But the longitudinal contribution 
is not completely lost, since it still appears
in $F_2^D$ (see eq.(\ref{numeric1}))

We insert the formfactor $1/(l_t^2+Q_0^2)$ which belongs to $\cal{F}$ into
eqs.(\ref{diff5}), (\ref{diff6}) and (\ref{diff8}) and perform the
integration over $l_t^2$ analytically. For convenience we introduce 
the variable $v$ which is defined as
$v=Q_0^2/(k_t^2+\alpha(1-\alpha)Q^2)=Q_0^2(1-\beta)/k_t^2$ for the quark
dipole and $v=Q_0^2/k^2=Q_0^2(1-z)/k_t^2$ for the gluon dipole.
$F_2^D$ is then presented in three separate
contributions ($F_2^D=F_a+F_b+F_c$) according to
eqs.(\ref{diff5})($F_a$), (\ref{diff6})($F_b$) and (\ref{diff8})($F_c$):
\beqn\label{numeric3}
F_a(\beta,Q^2,x_p)&=&\frac{1}{12 B_D Q_0^2}\;
\int_{4\beta Q_0^2/Q^2}^\infty
\frac{dv}{v^2\sqrt{1-4\frac{Q_0^2\beta}{Q^2v}}}
\;G^2(x_p,1/v)\;\frac{\beta}{x_p} \; (1-2\frac{Q_0^2 \beta}{Q^2 v})\;\;\cdot\\
&&\cdot\;\frac{1}{6}\,\frac{1}{1-\beta}\;
\left\{\,(1-2\beta)\ln\left(\frac{1}{\beta}\right)\;+\;
\left(\frac{1-2\beta+v}{\sqrt{v^2+2(1-2\beta)v+1}}-1+2\beta\right)\,\ln(v)
\nonumber\right.\\
&&\left.+\frac{1-2\beta+v}{\sqrt{v^2+2(1-2\beta)v+1}}\,\ln\left(\frac
{\sqrt{v^2+2(1-2\beta)v+1}\;-(1-2\beta+v)}{\sqrt{v^2+2(1-2\beta)v+1}\;
+(1-2\beta)v+1}\right)\right\}^2 \nonumber\;\;,
\eeqn
\beqn\label{numeric4}
F_b(\beta,Q^2,x_p)&=&\frac{1}{6 B_D Q^2}\;
\int_{4\beta Q_0^2/Q^2}^\infty
\frac{dv}{v^3\sqrt{1-4\frac{Q_0^2\beta}{Q^2v}}}
\;G^2(x_p,1/v)\;\frac{\beta^3}{x_p} \;\;\cdot\\
&&\cdot\;\frac{4}{3}\;\left\{\,\ln\left(\frac{1}{\beta}\right)\;+\;
\left(\frac{1}{\sqrt{v^2+2(1-2\beta)v+1}}-1\right)\,\ln(v)
\nonumber\right.\\
&&\left.+\frac{1}{\sqrt{v^2+2(1-2\beta)v+1}}\,\ln\left(\frac
{\sqrt{v^2+2(1-2\beta)v+1}\;-(1-2\beta+v)}{\sqrt{v^2+2(1-2\beta)v+1}\;
+(1-2\beta)v+1}\right)\right\}^2\nonumber
\eeqn
and
\beqn\label{numeric5}
F_c(\beta,Q^2,x_p)&=&\frac{1}{12 B_D Q_0^2}\;
\int_{Q_0^2/Q^2}^\infty \frac{dv}{v^2}\;G^2(x_p,1/v)\;\frac{\beta}{x_p}
\;\;\cdot\\
&&\cdot\;\frac{\alpha_s}{8\pi}\,\ln\left(\frac{v Q^2}{Q_0^2}\right)
\;\int_{\beta}^1\frac{dz}{z^2}\;\left[\left(1-\frac{\beta}{z}\right)^2\,+\,
\left(\frac{\beta}{z}\right)^2\right]\;\;\cdot\nonumber\\
&&\cdot\;\frac{9}{4}\frac{1}{(1-z)^2}\;\left\{[1+v-2z(1-z)]\,
\ln\left(\frac{1}{z}\right) \;+\;\ln(v)\;\;\cdot\right.\nonumber\\
&&\hspace{0.4cm}\cdot\;\left[v-1+2z(1-z)+\sqrt{v^2+2(1-2z)v+1}\,-\,
\frac{2z(1-z)}{\sqrt{v^2+2(1-2z)v+1}}\right] \nonumber\\
&&\hspace{0.3cm}+\left[\sqrt{v^2+2(1-2z)v+1}\,-\,\frac{2z(1-z)}
{\sqrt{v^2+2(1-2z)v+1}} \right] \;\;\cdot\nonumber\\
&&\hspace{3cm}\left.\cdot\;\ln\left(\frac{\sqrt{v^2+2(1-2z)v+1}-(1-2z+v)}
{\sqrt{v^2+2(1-2z)v+1}+(1-2z)v+1}\right)\right\}^2 \nonumber\;\;. 
\eeqn

Since the measurement is not performed at $t=0$ we have assumed a simple
exponential behavior in t, $\exp(B_D t)$, with the slope parameter $B_D$
taken from experiment \cite{LPS} ($B_D=5.9/GeV^2$). The integration over
t leads to the extra factor $1/B_D$. For $\alpha_s$ we
estimate a value of 0.25 which is a reasonable estimation for scales around $2-3
GeV$. The function $G$ is defined in eq.(\ref{f2_4}) and enters
the equations above without changing the parameters.
\begin{figure}[h]
  \begin{center}
    \leavevmode
     \epsfig{file=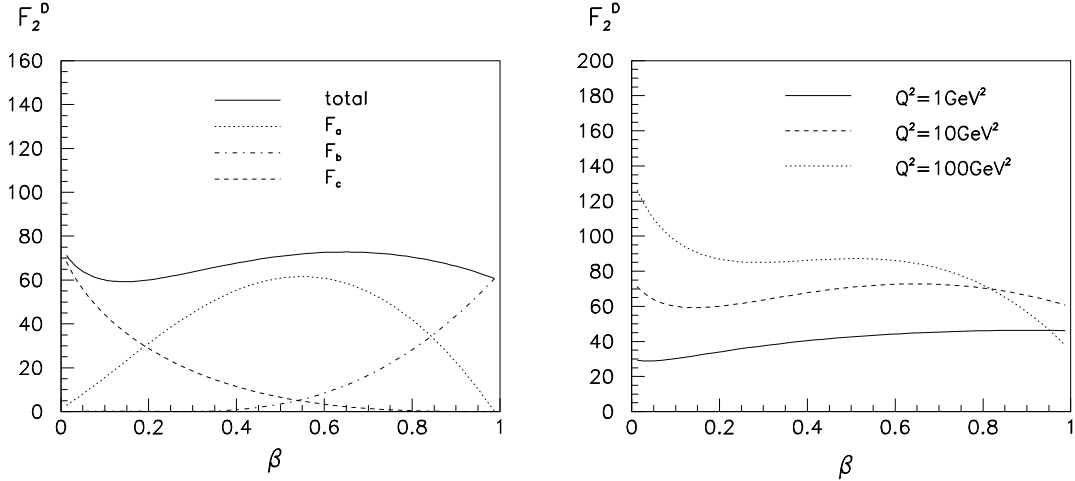,height=7cm}
  \end{center}
  \caption{$\beta$-spectrum.}
  \label{fig4}
\end{figure}
In fig.\ref{fig4} (first plot) we show the $\beta$-distribution for fixed
$x_{\fP}=5.0 \cdot 10^{-4}$ and $Q^2=10GeV^2$ with separate curves for
each of the three contributions $F_a, F_b$ and $F_c$. As was already
argued analytically we find three distinct regimes in the
$\beta$-spectrum: i) small $\beta$ where the configuration with a gluon in
the final state dominates ($F_c$), ii) medium $\beta$ where the exclusive
quark-antiquark production with transverse polarization is dominant
($F_a$) and iii) large $\beta$ where the longitudinal production 
of quark-antiquark pairs takes over ($F_b$). The second plot displays the change
in the shape of the $\beta$-distribution with increasing $Q^2$. It is
rather flat around $Q^2=10GeV^2$ before it starts tilting
when $Q^2$ is further increased (higher twist suppression at the large end and
logarithmic enhancement at the low end of $\beta$). 

The next figure (fig.\ref{fig5}) shows the $x_{\fP}$-distribution for
fixed $\beta$ and $Q^2$ (values as indicated in each graph). 
\begin{figure}[t]
  \begin{center}
    \leavevmode
     \epsfig{file=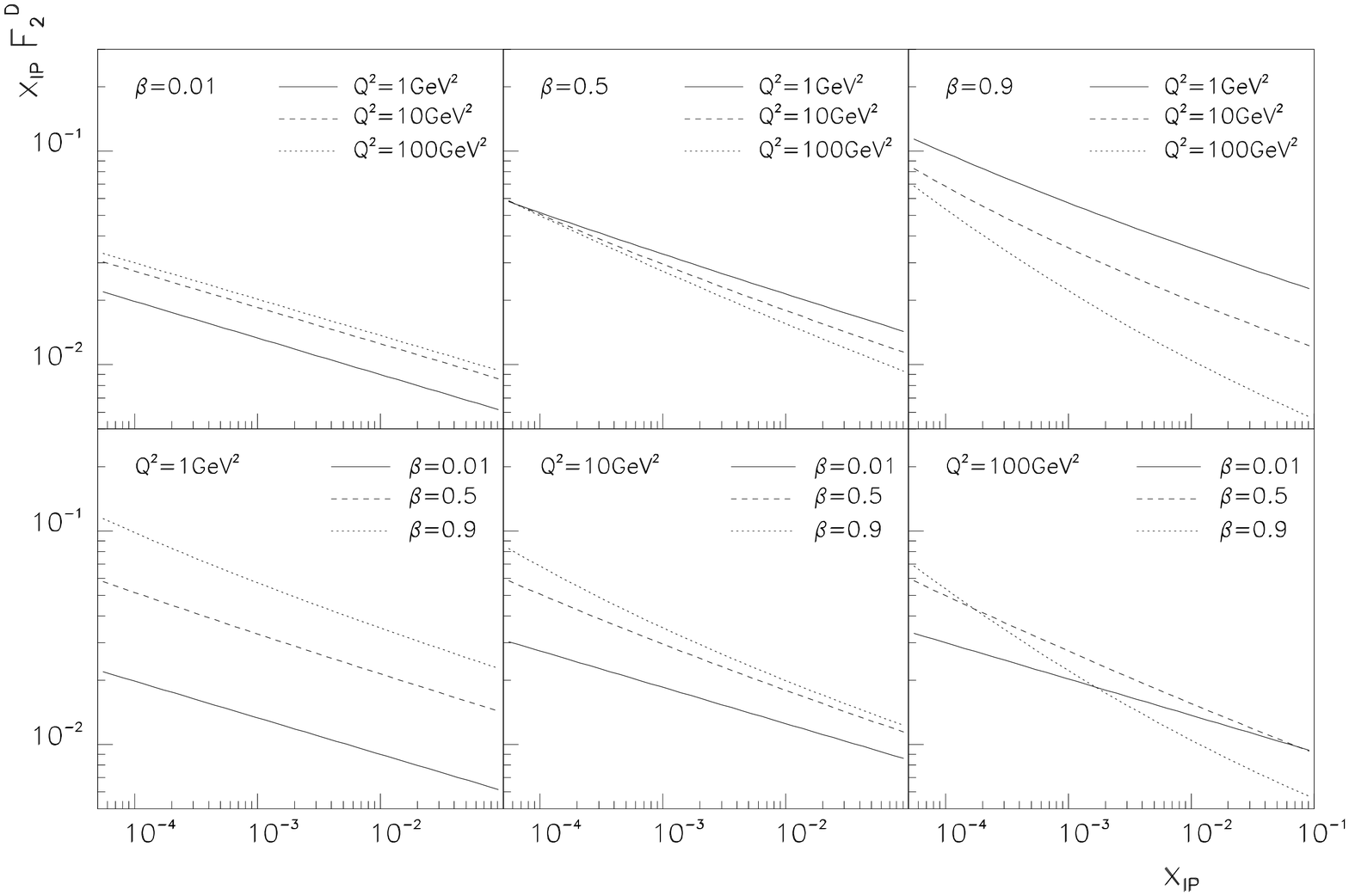,height=9.5cm}
  \end{center}
  \caption{$x_{\fP}$-distribution.}
  \label{fig5}
\end{figure}
The first row of plots starts at low $\beta$ with rather flat distributions 
in $x_{\fP}$ (the slope is 0.17 due to soft contributions, 
$\alpha_{\fP}=1.085$). 
The distributions become slightly more curved and steeper when
we move to larger $\beta$.
The change of the shape with $\beta$ results from the nonfactorizing ansatz 
where the Pomeron intercept depends on an intermediated (variable)
scale related to the size of the dipole ($Q_0^2/v$ in
eqs.(\ref{numeric3}), (\ref{numeric4}) and (\ref{numeric5})). 
At low $\beta$ this scale is small (large dipole)
and the Pomeron is dominantly soft. At high $\beta$ and in particular when the
higher twist (longitudinal) part takes over the intermediate scale
is, in average, rather hard (approximately $Q^2/4$) which leads to 
steeper distributions. This effect can roughly be interpreted as 
large mass states having a larger radius than small mass states.  
The curvature in the double log
plots is due to smearing when the scale is integrated. 
In the second row we see a similar behavior by changing $Q^2$. At small
$Q^2$ the intercept is frozen at a low (soft) value close to 1.085. 
There is only a little, barely visible effect due to smearing.
When $Q^2$ is increased the intermediate scale is pulled up 
and one finds again a steeper and slightly
curved distribution. In total we note that Regge-type factorization is
violated, i.e. the Pomeron intercept depends on $\beta$ and $Q^2$.

The scaling properties in diffraction are of special interest 
because they provide direct information about the Pomeron structure.
\begin{figure}[t]
  \begin{center}
    \leavevmode
     \epsfig{file=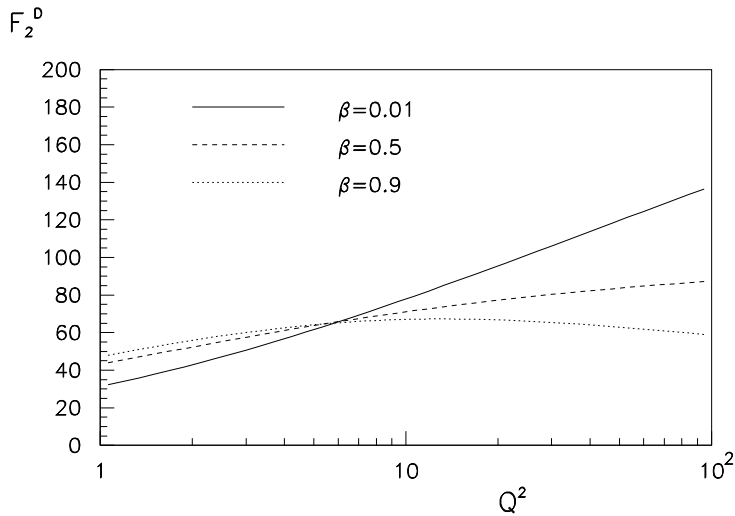,height=7.5cm}
  \end{center}
  \caption{$Q^2$-distribution.}
  \label{fig6}
\end{figure}
Fig.\ref{fig6} shows the dependence of $F_2^D$ on $Q^2$ for three different
values of $\beta$ (0.01, 0.
5 and 0.9) which as we know from fig.\ref{fig4} 
also distinguish between the three contributions $F_c, F_a$ and $F_b$.
(For very low $Q^2$ (close to 1GeV$^2$) our formula for $F_c$ 
has to be taken with care, since it is only 
computed to leading log($Q^2$) accuracy.)
The value for $x_{\fP}$ is again fixed at $5.0 \cdot 10^{-4}$.

The prediction for the slope in $Q^2$ is according to usual $Q^2$-evolution 
negative at large $\beta$. Fig.\ref{fig6}, however, shows a rise at low $Q^2$ 
for any value of $\beta$, even for the longitudinal contribution ($\beta=0.9$).
The latter
develops a maximum around $Q^2=10GeV^2$ before the asymptotic regime
is reached and the $1/Q^2$ (higher twist) suppression sets in.
For $\beta=0.5$ we also see an increase with $Q^2$ which then flattens
out towards a constant (leading twist scaling) behavior.
A rise over a certain range in the $Q^2$-distribution 
is not completely surprising, since $F_2^D$ vanishes when $Q^2$ 
approaches zero. What is surprising is the delay with which the
asymptotic (scaling) regime sets in. This effect seems to be model
dependent. Taking the hard Pomeron approach from ref.\cite{Pesch} as
example (plots can be found in ref.\cite{McDer}) the delay 
is even more pronounced. 
A precise measurement of the $Q^2$-scaling behavior seems to be a
promising tool to discriminate various Pomeron models.

A $Q^2$-scaling violation for rather large $\beta$ 
which persists far into the asymptotic region
can presumably not be reconciled with the 
dipole approach. An alternative scenario 
based on the hard component of the soft Pomeron would predict 
a log($Q^2$)-type behavior \cite{White}. 

We have not presented a comparison with data here. This can, however, be
found in \cite{H1diffwar}. The theoretical curve in \cite{H1diffwar} is based
on the same model as presented in this paper, only the values 
for the parameters have changed slightly 
with little impact on the $x_{\fP}$-spectra.  
\section{Summary}
We have derived the cross section for diffraction in Deep Inelastic
Scattering starting from two types of light cone wave functions, one
for a quark dipole (eq.(\ref{lw1})) and the second for a gluon dipole
(eq.(\ref{lw5})). The latter is of higher order in perturbation theory,
since a direct coupling of photons to gluons is lacking. We have
shown how the color dipole approach works for a multi parton
state provided one stays within the realm of leading twist and 
leading log($Q^2$) accuracy (strong ordering in impact parameter
space). Only the leading order quark-antiquark pair forms a dipole 
for any kinematics. The light cone wave function formalism was proven to
be consistent
with Feynman diagram calculations, but it should not be confused with the
general dipole approach of ref.\cite{Muedipole}. As soon as the strong
ordering in impact parameter is lost multiple dipoles may occur. The
wave function formalism presented here is not able to cope with this
configuration. 
  
The expression for the quark dipole (\ref{lw1}) is rather
well known, the second expression (\ref{lw5}) for the two gluon dipole is new. 
It is valid over the complete range of invariant mass of the two gluons
and therefore an extension of an earlier derived version which was
limited to large masses (triple Regge limit, eq.(\ref{lw5.5}))
  
As model for the Pomeron we have considered color zero two gluon exchange
which is easily generalized to multigluon exchange.
All gluons at each leg of the dipole merge into a single vertex, i.e.
there is only a single interaction point in impact parameter space.
We can factorize the dipole from the target ($k_t$-factorization scheme)
and parametrize all unknown contribution in terms of an
unintegrated gluon structure function. This factorization works for two
simple perturbative gluons, for shadowing corrections \cite{Lev} and even
for scattering in a nonperturbative classical field. It should be
possible to reformulate the results in ref.\cite{BuHe} along the line of
our dipole approach. In the semiclassical approach of ref.\cite{Heribert}
the gluon density is directly related to the unintegrated structure
function, and the Landshoff-Nachtmann model \cite{Diehl} can as well be
identified with an appropriate unintegrated gluon structure function.

In this paper, however, a more phenomenological ansatz was chosen. A
parametrization for the unintegrated structure function was established with
parameters determined from inclusive Deep Inelastic Scattering ($F_2$) and then
inserted into the corresponding expression for diffraction. An important feature
of our parametrization is the scale-dependence of the Pomeron intercept
which results in a variation of the $x_{\fP}$-slope.
The scale is roughly the inverse size of the dipole (not $Q^2$) 
which has to be integrated over. Its average is close to the soft
scale of 1GeV, but increases slightly with $\beta$ and
$Q^2$ which then causes the $x_{\fP}$-distribution to become steeper.
The lower limit for the slope is given by the soft
Pomeron intercept. The $\beta$-spectrum is subdivided into three regions
each being dominated by the following contributions: i) the gluon dipole
at small $\beta$, ii) the quark dipole with transverse polarized photons 
at medium $\beta$ and iii) the quark dipole with longitudinal
polarized photons (higher twist) at large $\beta$.
Around $Q^2=10GeV^2$ the total spectrum is rather flat, but it
starts tilting when $Q^2$ is increased (falling from $\beta=0$ to
$\beta=1$). 

The $Q^2$-distribution is of special interest because it helps revealing
the structure of the Pomeron. We find that for $\beta > 0.5$ the slope
in $Q^2$ is positive up to $Q^2\sim$ 10GeV$^2$ and then flattens out
(leading twist) or turns down (higher twist at large $\beta$) (see
fig.\ref{fig6}). A comparison with data from H1 was performed in
ref.\cite{H1diffwar} where the agreement is found to be reasonable.
The main deviation between theory and data is due to  secondary
exchanges which have not been included in this paper. 
The LPS-data from ZEUS \cite{LPS} also seem to agree quite well.

So far only leading order and most important next-to-leading order
contributions have been taken into account. To obtain more precise
prediction for $\beta$- and $Q^2$-distributions one needs to perform a
complete $Q^2$-evolution which will be subject of another publication.
Also of interest is a next-to-leading order diffractive jet-analysis.
This requires, however, a full and consistent next-to-leading order calculation
which goes beyond the light cone wave function approach of this paper.  
\vspace{.5cm}\\
{\bf Acknowledgements:} I am very grateful to J.Bartels and A.White 
for valuable discussions and I thank the Argonne ZEUS-group for
providing information about data and experiment.
This work was supported by the US Department of Energy, Contract
W-31-109-ENG-38.

\end{document}